\documentclass[preprint,12pt]{elsarticle}

\usepackage{graphicx}
\usepackage{amssymb}
\usepackage{amsthm}

\journal{}

\begin{document}

\begin{frontmatter}

\title{Scaling analysis of time series of daily prices \\ from stock markets of transitional economies \\ in the Western Balkans}

\author[prvi]{Darko Sarvan}

\address[prvi, cetvrti]{Faculty of Physics, University of Belgrade, P.O. Box 368, 11001 Belgrade, Serbia}

\author[drugi]{Djordje Stratimirovi\'c}

\address[drugi]{Faculty of Dental Medicine, University of Belgrade, Dr Suboti\'ca 8, 11000 Belgrade, Serbia}

\author[treca]{Suzana Blesi\' c\corref{cor1}}
\ead{suzana@imi.bg.ac.rs}

\cortext[cor1]{Corresponding author}

\address[treca]{Institute for Medical Research, University of Belgrade, P.O. Box 102, 11129 Belgrade, Serbia}

\author[cetvrti]{Vladimir Miljkovi\' c}

\begin{abstract}
In this paper we have analyzed scaling properties of time series of stock market indices (SMIs) of developing economies of Western Balkans, and have compared the results we have obtained with the results from more developed economies. We have used three different techniques of data analysis to obtain and verify our findings: Detrended Fluctuation Analysis (DFA) method, Detrended Moving Average (DMA) method, and Wavelet Transformation (WT) analysis.
We have found scaling behavior in all SMI data sets that we have analyzed. The scaling of our SMI series changes from long-range correlated to slightly anti-correlated behavior with the change in growth or maturity of the economy the stock market is embedded in.
We also report the presence of effects of potential periodic-like influences on the SMI data that we have analyzed. One such influence is visible in all our SMI series, and appears at a period $T_{p}\approx 90$ days. We propose that the existence of various periodic-like influences on SMI data may partially explain the observed difference in types of correlated behavior of corresponding scaling functions.

 %    } % end of PACS codes
\end{abstract}


\end{frontmatter}

\section{Introduction}
\label{intro}
Analysis of scaling and fluctuation dynamics has been applied extensively outside of what is traditionally defined as a physical sciences domain \cite{ref1,ref2,ref3,ref4,ref5,KE05,SLK08,CFL09,FCET13,GGM12}. In the recent decades, new methods of statistical physics have been successfully utilized to analyze dynamics of price changes on stock markets, which has led to a number of interesting new findings and explanations of market behavior \cite{ref6,ref7,ref8,ref9}. Economic systems remain among the most interesting complex systems open to investigation by physicists \cite{ref44}.\\
Following extensive research in the area of econophysics \cite{ref15} of national and international stock markets \cite{ref10,ref11,ref12,ref13}, we were interested to contribute to this body of knowledge by analyzing the dynamics of market behavior of transitional economies in the Western Balkans, and to compare data from these developing economies with data from more economically developed countries. Analyzes of stock market behavior of the emerging economies of South America \cite{ref13''}, or the developing Asian or African markets \cite{ref13'''} have shown that the values of scaling exponents, calculated from the time series of stock market indices, could be used to estimate the efficiency \cite{ref41} of markets in question. The purpose of this paper is to, by applying the theoretical approach of statistical physics, offer a new perspective to stock market dynamics in the Western Balkans, contribute to better understanding of the development processes in the region's  economies, and add to the overall understanding of forces that drive global markets.\\
In this paper we apply scaling analysis to time series made of daily closing values of stock market indices (SMIs). In doing so, we accept findings of previous studies about the existence of scaling in SMI data, and are interested in a kind of scaling behavior that the series we analyze exhibit. Namely, earlier studies \cite{ref14} have confirmed the existence of long-range type of correlated behavior in SMI data, identifiable through power-law type of behavior of power spectra, correlation and fluctuation functions that describe that data.  Recent studies \cite{ref16} have also indicated the possibility of both correlated and anticorrelated (persistent and non-persistent) long-range type behavior of SMI data, which can be connected to the degree of development of the stock market (and/or the economy that accommodates it) that is analyzed. We question this finding, using new data comprised of SMI time series from developing economies of transitional countries in the Western Balkans. We also examine these data series comparatively at both global and local time scales, in order to asses the specific influences (trends, or cycles) that bring about the observed types of correlation.\\
We have used three different techniques of data analysis, in order to assess and verify the nature of stochastic dynamics of the selected stock prices time series. First, we conducted the Detrended Fluctuation Analysis (DFA), to assess the dynamic properties of our time series. Then, we confirmed the DFA results with the Detrended Moving Average (DMA) method. Finally, we utilized the Wavelet Transform (WT) analysis on all data sets, so as to get an independent verification of the DFA and DMA results, and to highlight additional findings. In using WT we specifically rely on the possibility  \cite{ref18} to examine the existence of periodic or non-periodic cycles in data, if such trends are discerned by DFA or DMA method.

\section{Data}
\label{sec:1}
We have analyzed financial time series of nine stock market indices: four of the developing economies in the Western Balkans -- BELEXline index of stock market of Republic of Serbia, SASX10 index of stock market of Federation of Bosnia and Herzegovina, BIRS index of market of Bosnian Serbian entity Republic of Srpska, and a MONEX20 index of the Republic of Montenegro; three of emerging European economies -- CROBEX index of the stock market of Republic of Croatia, BUX index of the market of Republic of Hungary, and XU100 index of the Republic of Turkey; and two of developed economies -- DAX index of German economic market and the CAC40 index of the French market. All the analyzed indices are the all-share type.
Table 1. lists general characteristics of the time series we have analyzed.\\

\begin{table*}
\caption{General characteristics of the SMI time series analyzed in this paper.}
\label{tab:1}
\begin{tabular}{llllll}
\hline\noalign{\smallskip}
SMI name (economy) & Recording period & Total days $N$\\
\noalign{\smallskip}\hline\noalign{\smallskip}
BELEXline (Serbia) & October 1, 2004 - December 30, 2011 & 1828\\
BIRS (Republic of Srpska) & April 27, 2004 - December 23, 2011 & 1839\\
%BIRS, all-share index (Bosnia and Herzegovina -- Republic of Srpska) & April 27, 2004 - December 23, 2011 & 1839\\
MONTEX20 (Montenegro) & January 9, 2004 - October 18, 2013 & 2423\\
SASX10 (Bosnia and Herzegovina) & April 28, 2003 - October 21, 2013 & 1926\\
CROBEX (Croatia) & April 1, 2000 - December 22, 2011 & 2991\\
BUX (Hungary) & April 1, 1997 - December 22, 2011 & 3679\\
XU100 (Turkey) & January 4, 1988 - October 20, 2013 & 6441\\
DAX (Germany) & November 26, 1990 - December 30, 2011 & 5338\\
CAC40 (France) & March 1, 1990 - December 30, 2011 & 5526\\
\noalign{\smallskip}\hline
\end{tabular}
\end{table*}

In order to assess scaling properties of SMI time series in question, we analyzed the series of logarithmic returns $R(t)$ of market indices $S(t)$, calculated \cite{ref7} as a difference of logarithmic values of market indices at time $t$:
$R(t) = logS(t+\Delta t)-logS(t)=log(S(t+\Delta t)/S(t))$,
where the lag period $\Delta t$ is one day, a time interval of recording of index values $S(t)$. All of the analyzed time series of prices on the stock markets $S(t)$ are publicly available (from official Web-sites of markets in question), and are given in local currencies. The time series used are, depending mainly on the market development level, of varying duration (see Table 1.).\\

\section{Detrending methods for scaling analysis}
\label{sec:2}
In this paper we have utilized two detrending scaling methods, commonly used for the detection of correlations within data time series.\\
We have first used a Detrended Fluctuation Analysis (DFA) technique. The variant of the standard DFA method that we have used \cite{ref2} requires three consecutive steps. In the first step, for each sequence of $R(k)$, a partial sum, or the integrated series, is calculated:
$y(l)=\sum_{k=1}^l[R(k)-R_{ave}]\>,$ where $R_{ave}$ is the average stock market logarithmic return value, that is, $R_{ave}={1\over N}\sum_{k=1}^NR(k)$, with $N$ being the total number of recorded values for a given series (the SMI number $k$ plays the role of time $t\,$). In the next step, the entire series of $y(l)$ is divided  into a set of overlapping segments \cite{ref19} of the length $n$ and a local trend for each segment is calculated. Local trend is the linear or polynomial least-squares fit for the segment data. The order of polynomial that defines the local trend represents the order of DFA method \cite{ref20}. In this paper we use DFA functions of the second order. For the new series, one has to define the so-called detrended walk $y_{n,i}(l)$ which is, for a given segment, the difference between original series of partial sums $y(l)$ and the local trend. Finally, one has to calculate the variance about the local trend for each segment and determine the average of these variances over all segments, which brings about the detrended fluctuation function
$$F(n)=\sqrt{{1\over{(N-n+1)n}}\sum_{i=1}^{N-n+1}\sum_{l=1}^l(y_{n,i}(l))^2}\>.
\eqno (1)$$\\
By increasing the segment length $n$ the function $F(n)$ increases as well. When the analyzed time series follows a scaling law, the DFA function is of a power-law type, that is, $F(n)\propto\ n^{\alpha}$, with $0 \leq \alpha \leq 1$. In the case of short-range data correlations (or no correlations at all), the detrended walk displays properties of a standard random walk (white noise), while $F(n)$ behaves as $n^{1/2}$ \cite{ref2}. For data with power-law long-range correlations one may expect that $\alpha>0.5$, while in long-range anti-correlated case we have $\alpha<0.5$. When scaling exists, the exponent $\alpha$, associated with the detrended fluctuation function $F(n)$, can be related \cite{ref21} to the Fourier power spectrum exponent $\beta$ through the scaling relation $\alpha=(\beta+1)/2$.

For our study we use the Detrended Moving Average (DMA) technique next. It has recently been proposed \cite{ref22} that the standard DFA method could be verified using the DMA technique, particularly for lower values of scale. The original DMA method was introduced \cite{ref23} in order to remove a drawback of the standard DFA technique, the fact that the curve made of DFA local trends is discontinuous at positions $i=nk; k = 1, 2, . . . $. This may lead to false determination of the value of scaling exponent $\alpha$, particularly at small scales $n$, or in short data sets \cite{ref22}.\\
The variation of a standard DMA method that we use here, a Centered Detrended Moving Average (cDMA) scaling technique \cite{ref24}, uses the moving average to remove trends in data by defining the detrended curve at segment size $n$ as:
$$ y_{n,i}(l)= x(i) - \frac{1}{n}\sum_{j=-(n-1)/2}^{(n-1)/2} x(i+j) \>, \eqno (2)$$
\noindent while the remaining steps in the calculation procedure, as well as the definition of the detrended fluctuation function, usually denoted as $\sigma(n)$ in cDMA case, are the same as in the DFA method and in Eq.(1). The advantages of the cDMA algorithm are attributed to the better low-pass performance of the moving average compared to polynomial filters \cite{ref25}. The calculated scaling exponent, usually identified as a Hurst exponent $H$, can be directly compared to the DFA exponent $\alpha$.

\section{Wavelet transformation method}
\label{sec:3}

We have applied the Wavelet Transformation (WT) method to analyze time series of stock market indices logarithmic returns in order to verify the DFA and cDMA results through an independent method, and gain a new insight into the findings of detrending methods.\\
The continuous wavelet transform \cite{ref262,ref263} of a discrete sequence $R(k)$ is defined as the convolution of $R(k)$ with wavelet functions $\psi_{a,b}(k)$ in the following way:
$$W_P(a,b)=\sum_{k=0}^{N-1}R(k)\psi^*_{a,b}{(k)}\>, \eqno (3)$$
\noindent with $a$ and $b$ being the scale and translation-in-time (coordinate) parameters, and $N$ the total length of the SMI series studied. The wavelet functions $\psi_{a,b}(k)$, used in Eq.(3), are related to the scaled and translated version of the mother--wavelet $\psi_0(k)$, through
$$\psi_{a,b}(k)={1\over \sqrt a}\,\psi_0\biggl({{k-b}\over{a}}\biggr)\>. \eqno
(4)$$\\
In this paper, we find it convenient to use the standard set \cite{ref27} of derivatives of Gaussian (DOG) wavelet functions.\\
In order to obtain the kind of results comparable with those of the DFA and cDMA method, we have calculated the so--called scalegrams (mean wavelet power spectra) $E_W(a)$, that are defined by $$E_W(a)=\int
W^{2}(a,b)db\>. \eqno (5)$$\\
The scalegram $E_W(a)$ can be related \cite{ref28} to the corresponding Fourier power spectrum $E_F(\omega)$ via the formula
$$E_W(a)=\int
E_F(\omega)|\hat{\psi}(a\omega)|^2d\omega\>. \eqno
(6)$$\\
This formula implies that if the two spectra, $E_W(a)$ and $E_F(\omega)$, exhibit power-law behavior, then they should have the same power-law exponent $\beta$, comparable to scaling exponent(s) $\alpha$ and $H$ through scaling relation $\alpha=(\beta+1)/2$.

\section{Scaling properties of SMI time series}
\label{sec:4}

Our results confirm the findings of previous studies on the existence of scaling in SMI time series \cite{ref6,ref7,ref8,ref9,ref10,ref11,ref12,ref13}.  In all of the analyzed cases we have found power-law type behavior of DFA and cDMA functions, and of the WT power spectra, and have calculated scaling exponents $\alpha$, $H$, and $\beta$ from log-log graphs of these functions. An example of a typical result we have obtained for all our data is given in Fig.1, where an except of a DAX time series is depicted, together with the calculated DFA, DMA, and WT functions for this data series. In Fig.1, the WT function is re-calculated so that its slope can be directly compared to the one of DFA and DMA functions, following the scaling equation $\alpha=(\beta+1)/2$.

We have also found differences in scaling properties of the analyzed series. In Fig.2, combined results of the DFA, cDMA, and WT methods are repeated. The columns of Fig.2 give graphs of obtained DFA, cDMA, and WT functions, as functions of scale, respectively, while the rows correspond to the illustrative cases of three analyzed stock market categories - Serbian BELEXline index, representing the transitional markets of Western Balkans, Croatian CROBEX index, depicting the emerging European markets, and the CAC40 index of a French stock market, representing the markets of developed world economies. The DFA function used is of the second order (denoted DFA2), and wavelet basis is made of a DOG (derivatives of Gaussian) wavelets of the first order (denoted DOG1). Graphs in Fig.2 depict only the range of scales where the straight lines were fit to scaling functions, i.e. only the small scales range. We have chosen to restrict our study of scaling properties of SMI series to this range of scales for, having in mind the average length of the SMI series analyzed (with N of the order of ${10}^3$), we could expect to obtain statistically relevant results within this range \cite{ref21,ref22}. In addition, in choosing the fitting range in each particular case, we considered the behavior of all three scaling functions (e.g., some of our functions have crossovers, probably due to the preparation of data \cite{ref282}, that also limit the range of available scales). Our findings are listed in Table 2.

The scaling behavior of DFA, cDMA, and WT functions points to the existence of correlations within the SMI time series. The  values of scaling exponents, calculated as slopes of corresponding scaling functions, quantify the difference in types of correlated behavior of SMI series of different economies. Namely, the values of scaling exponents decline with the level of development of analyzed stock markets, decreasing from $\alpha, H>0.5$ or $\beta>0$, and crossing the $\alpha, H=0.5$ or $\beta=0$ line, as we go from developing new economies of Western Balkans to developed economies of Europe. We thus see a shift from pronounced long-range correlated behavior, found in BELEXline, BIRS and MONTEX20 series, through slightly long-range correlated behavior found in SASX10, CROBEX, BUX and XU100 series, to uncorrelated or slightly long-range anti-correlated behavior found in DAX and CAC40 indices series (see Fig.2).\\
Values of calculated scaling exponents for all the techniques of data analysis used, and all the data analyzed, are given in Table 2. In order to ease the comparison of obtained results, Table 2. gives, along with listed obtained values of WT exponents $\beta$, their corresponding calculated (through scaling equation) value of $\alpha (H)$. For our data, obtained DFA and cDMA exponents are generally similar, and slightly underestimate the values obtained from the WT analysis. Our DFA and cDMA results do not differ profoundly, as was reported before \cite{ref22}, in the very small scales ($n<10$) regions. This region was omitted from our analysis; if necessary to consider, we would, following Bashan et al. \cite{ref22}, rely on the obtained WT values as accurate in the very small scales region, in which case the scales could go as low as $n\approx 5$.

\begin{table}
\caption{Values of scaling exponents calculated by use of the DFA, DMA, and WT methods, for all the SMI data analyzed.}
\label{tab:2}
\begin{tabular}{llllll}
\hline\noalign{\smallskip}
SMI & Fitting Range & $\alpha$ & $H$ & $\langle H \rangle$ & ($\beta$) $\beta_{\alpha / H}$\\
%SMI & Fitting Scale Range & $\alpha$ & $H$ & $\langle H \rangle$ & $\beta_{\alpha / H}$\\
\noalign{\smallskip}\hline\noalign{\smallskip}
BELEXline & 10-170 & 0.72 & 0.73 & 0.73 & (0.46) 0.73\\
BIRS & 10-230 & 0.69 & 0.69 & 0.69 & (0.36) 0.68\\
MONTEX20 & 10-240 & 0.66 & 0.65 & 0.62 & (0.30) 0.65\\
SASX10 & 10-200 & 0.56 & 0.55 & 0.54 & (0.12) 0.56\\
CROBEX & 10-300 & 0.52 & 0.56 & 0.55 & (0.18) 0.59\\
BUX & 10-500 & 0.55 & 0.54 & 0.53 & (0.10) 0.55\\
XU100 & 10-500 & 0.53 & 0.54 & 0.52 & (0.10) 0.55\\
DAX & 10-500 & 0.45 & 0.47 & 0.46 & (0) 0.50\\
CAC40 & 10-500 & 0.44 & 0.48 & 0.46 & (-0.04) 0.48\\
\noalign{\smallskip}\hline
\end{tabular}

\end{table}

\section{Trends in scaling: cycles in SMI data}
\label{sec:5}

In all the SMI data sets that we analyzed, we noticed indicators of influence of the periodic-like trends within the analyzed data. As reported before \cite{ref30}, if a time series has an embedded periodical trend, it will emerge in a specific way in the behavior of scaling functions - for detrending methods, the scaling function will display existence of three crossover points, whose positions depend solely on the period $T$ of the embedded trend. Our DFA and cDMA functions do, indeed, show this kind of behavior. The regions of influence of periodic-like trends can already be seen in the graphs of DFA and cDMA functions, given in the left and central column of Fig.2, as an increase of the slopes of the scaling functions, followed by an immediate decrease, to the same as or to a different level than on onset. The existence of periodic-like trends, in all shown cases (and in all our data), is also seen in the right column in Fig.2, as a protrusion in WT functions, at scales $n\approx 90$. Given that, in our case, the unit of scale equals a time unit of one day, this gives us a time period of $T\approx 90$ days.\\
As this is, to our knowledge, a new finding, we have made an effort to be precise in description of our observations. We have thus employed the WT method to clarify these results. The advantage of the WT method in detecting trends in the data, particularly periodic trends, lies in the ability to utilize different wavelet functions as a basis for the wavelet transform. In that manner, one is able to find and choose a set of functions that closely follow the analyzed signal \cite{ref18,ref27}, and avoid ambiguous results. In our case, different orders of the used DOG wavelet can be utilized to emphasize and investigate possible periodic features of the SMI data. Results of this approach are given in Fig.3, where WT power spectra of the first (DOG1) and the tenth (DOG10) order are depicted, for the three representative SMI series (BELEXline, CROBEX, and CAC40). Application of wavelets of a higher order has indeed magnified the alterations in scaling behavior of WT functions. One of these alterations is visible in all our data (see Fig.3); positions of centers of this change are at $T_{p}\approx 90$ days. For the SMI series of the developed world economies (DAX, CAC40), this is the only visible effect of the use of DOG10 (in comparison to DOG1), while in the time series of SMIs of emerging and transitional European economies one can see two (XU100, BUX, CROBEX) or even three (BELEXline, BIRS, MONEX20, SASX10) such effects, in the regions of scales bellow $T_{p}$ (see Fig.3).\\

\section{Time-dependent SMI data analysis}
\label{sec:6}

In order to gain another insight into the local complexity of our SMI data, we have applied the time-dependent DMA (tdDMA) algorithm to all  SMI series. We apply the tdDMA algorithm \cite{ref25} on the subset of data in the intersection of the SMI signal and a sliding window of size $N_{s}$, which moves along the series with step $\delta_{s}$. The scaling exponent $H$ is calculated for each subset, according to the cDMA procedure described above, and a sequence of local, time-dependent Hurst exponent values is obtained. The minimum size of each subset $N_{min}$ is defined by the condition that the scaling law $\sigma(n)\propto\ n^{H}$ holds in the subset, while the accuracy of the technique is achieved with appropriate choice of $N_{min}$ and $\delta_{min}$ \cite{ref26}.  We have chosen windows of up to $N_{s} = 1000$, with the step $\delta_{s} = 2$ for our tdDMA algorithm, while the scaling features are studied in the region $n\in[2,500]$. Results obtained for the three representative cases (BELEXline, CROBEX and CAC40) are given in Fig.4. \\
We have confirmed (see Fig. 4) a previous finding \cite{ref25} that, in general, the financial time series display a visible local variability of the scaling exponent $H$, providing evidence that a complex evolution dynamics characterize the financial returns $R(t)$. Our data show the (local) variability of the exponent $H$ around it's calculated average (global) value (shown as a horizontal line in graphs of Fig.4 and listed, for all the SMI series, in Table 2.), which is not significantly dependent on the level of development of the analyzed stock market. Our results do not show any discernible periodic effects to SMI data, probably indicating that these effects, if exist, are of a more complex nature \cite{ref32}.\\

\section{Discussion and conclusions}
\label{sec:7}

In this paper we have analyzed scaling properties of time series of stock market indices (SMIs) of developing economies of Western Balkans, and have compared the results we have obtained with the results from more developed economies. We have used three different well established techniques of data analysis to obtain and verify our findings: Detrended Fluctuation Analysis (DFA) method, Detrended Moving Average (DMA) method, and Wavelet Transformation (WT) analysis.

We have found scaling behavior in all SMI data sets that we have analyzed. We have also observed that the nature of long-range correlations in our data sets depends on the level of development of market economy in question. Namely, scaling of SMI series changes from long-range correlated to slightly anti-correlated behavior, i.e. the appropriate scaling exponents decrease in value with the increase in growth and/or maturity of the economy the stock market is embedded in. Scaling exponents $\alpha$, $H$, and $\beta$, corresponding to the DFA, DMA, and WT technique, all cross the $0.5$ (and zero) line, marking this alteration.

Similar results have been reported before \cite{ref16,ref34,ref35}. The clear differentiation of scaling behavior connected to the level of development of the economy is reproduced and confirmed here in the case of the economies of the Western Balkans. Scaling functions of these developing economies all exhibit a long-range correlated behavior in small-scale regions. We find the economies of the Western Balkans to have $\alpha, H > 0.5$ ($\beta > 0$), opposite to developed world economies, which have $\alpha, H \leq 0.5$ ($\beta \leq 0)$. The observed sensitivity of scaling exponents to the level of development of economies, as well as the difference of the observed stock market indices behavior to the ideal uncorrelated case, could be related to a higher degree of possibility for arbitrage and profit in the underdeveloped stock markets \cite{ref36}.

We also report the presence of effects of potential periodic-like influences, or cycles \cite{ref362} in SMI data. One such influence is visible in all our SMI series, irrespective of the level of development of the analyzed stock market. It appears at a period $T_{p}\approx 90$ days, in all the series analyzed. This may be a sign of influence of a global economic factor that could be related to the period of release of various quarterly financial reports, both in Europe and in the United States \cite{ref37,ref38}. This, and other periodic-like influences found in time series of developing and emerging economies, particularly in economies of the Western Balkans, remain open to further investigation and discussion. These influences, if correctly detected, would also help explain the observed dynamics of Balkan stock markets, and the distinction between these developing markets and other developed economies \cite{ref39,ref40}.

From the data analysis point of view, the existence of various periodic influences on SMI data may also partially explain the observed difference in types of correlated behavior of corresponding scaling functions. As one can notice in Fig.3, the presence of periodic-like trends in scaling of DOG10 WT functions results in a rise of slopes of the matching DOG1 curves (and, consequently, of comparable DFA and cDMA functions), bringing about slightly or highly correlated behavior of stock markets of emerging or developing European economies. The application of time-dependent scaling analysis (tdDMA) proved that these influences are of a complex type, that is, they can not be easily distinguished from a local correlations profile. Therefore, the exact nature of the observed behavior remains to be studied, opening several directions for future research. Using the techniques utilized here, or any of their variants, long-range correlated SMI series could be analyzed time-dependently, aiming for better understanding of the observed behavior, and, ultimately, to model and reduce uncertainty and to assist in stock market development of new economies.

\noindent\textbf{Acknowledgment:} This work was supported by Serbian Ministry of Education and Science Research Grant no. 171015.

%\bibliographystyle{plainnat}
%\bibliographystyle{epj}
%\bibliography{bib-database}

\begin{figure}[fig1]
\resizebox{0.78\hsize}{!}{\includegraphics*{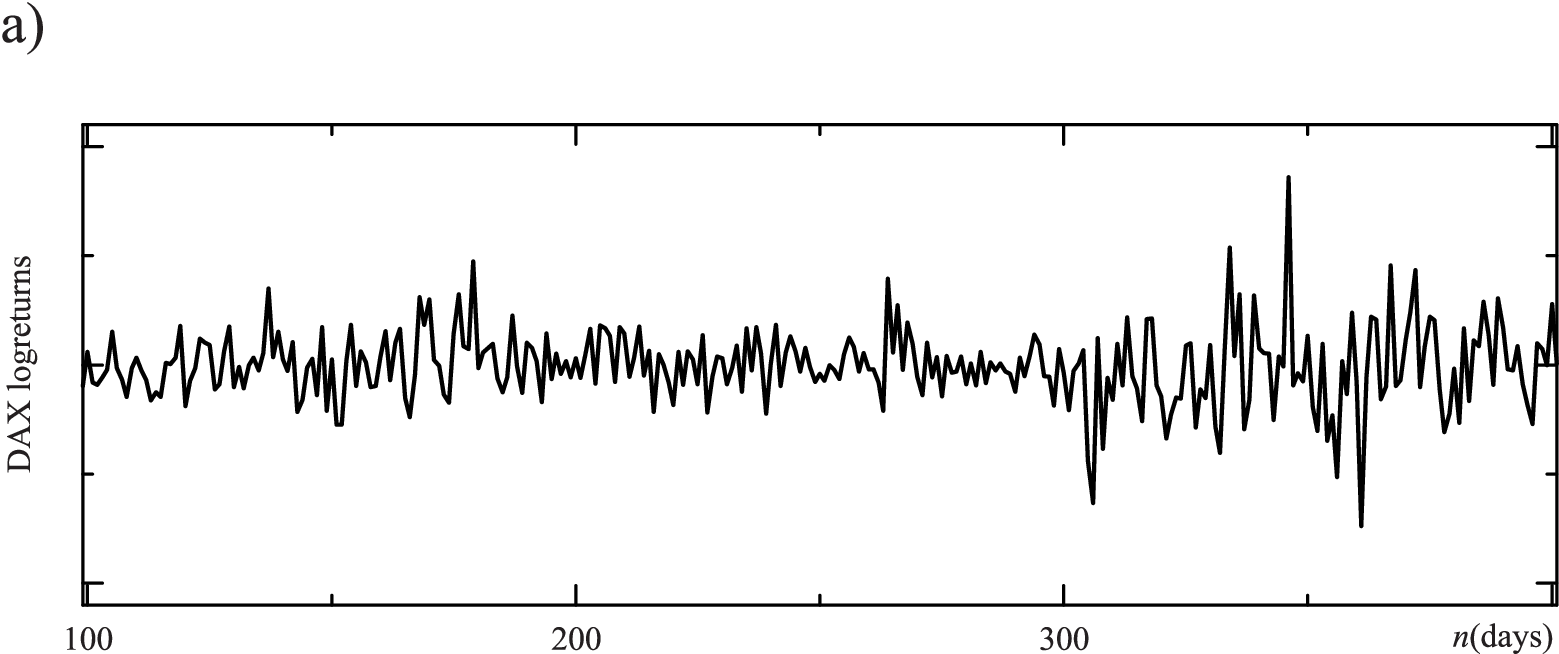}}
{\vskip 0.5cm}
\resizebox{0.78\hsize}{!}{\includegraphics{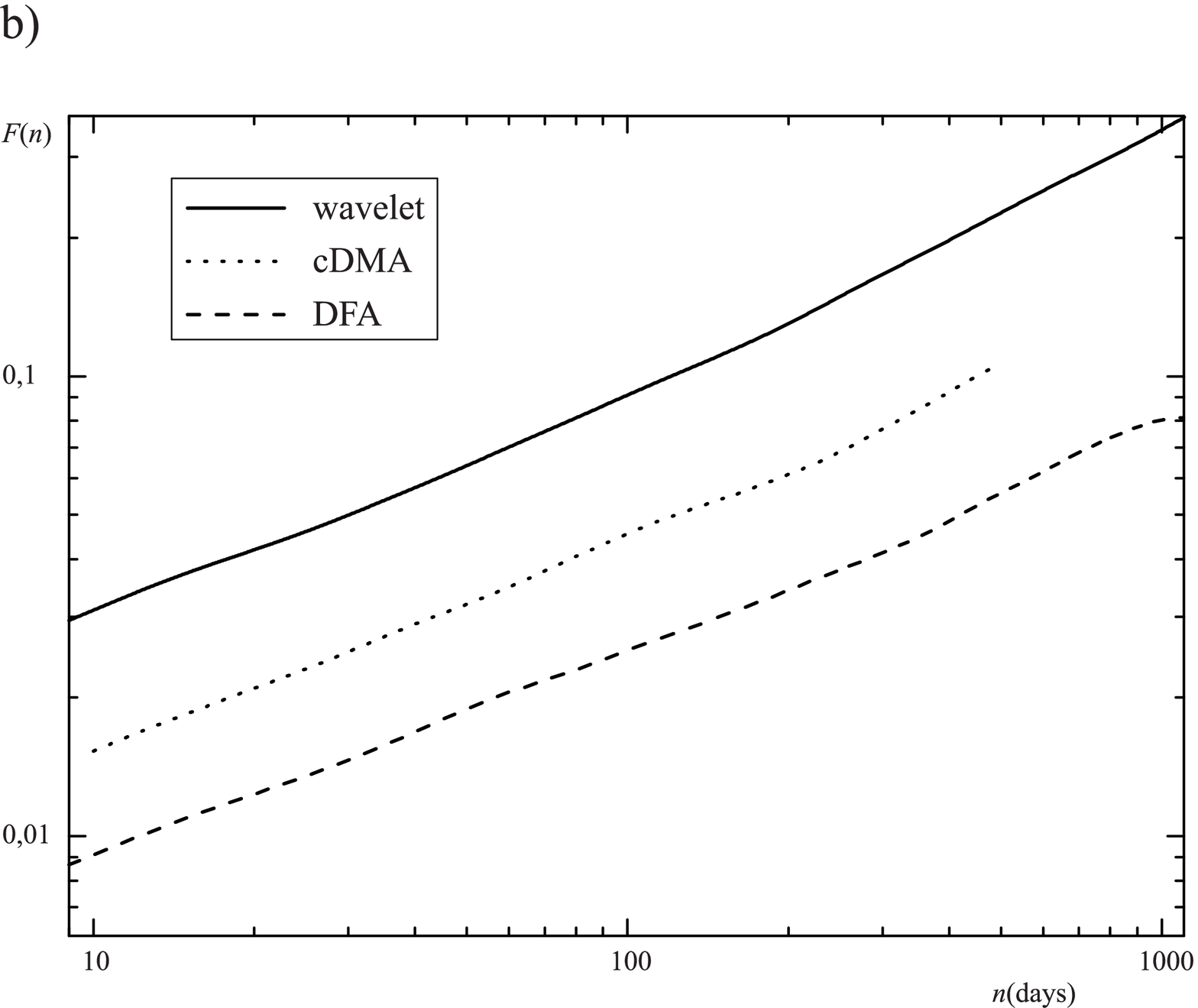}}
\caption{An example of DAX SMI series that demonstrates presence of scaling in data: a) an except of DAX time series, and b) the calculated behavior of DFA, DMA, and WT functions for DAX data. The WT function is re-calculated so that its slope can be directly compared to the one of DFA and DMA functions, following the scaling equation $\alpha=(\beta+1)/2$.}
\label{fig1}
\end{figure}

\begin{figure*}[fig2]
%\centering
\includegraphics[scale=0.78]{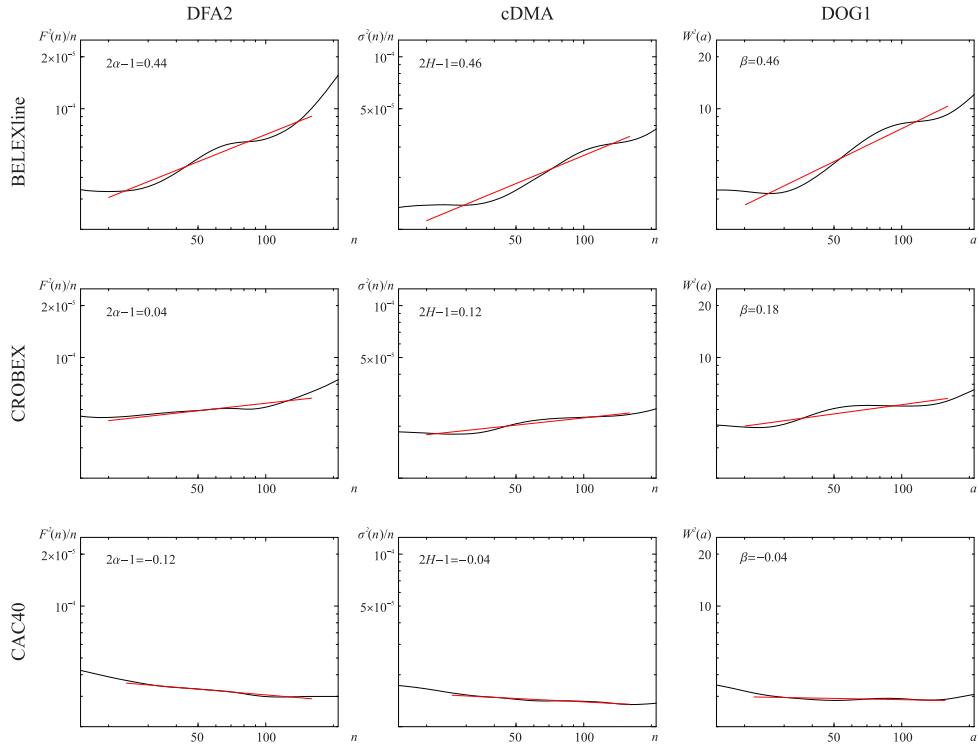}
%\resizebox{0.78\hsize}{!}{\includegraphics*{graph2a.eps}}
\caption{A three-part figure that demonstrates the presence of different types of scaling in SMI data. The columns give DFA2, cDMA, and DOG1 functions (as functions of scale), while the rows correspond to the illustrative SMI cases of three analyzed market categories - BELEXline index, CROBEX index, and the CAC40 index. The DFA and cDMA functions of the left and central column are depicted in the form $F^{2}(n)/n$ versus $n$, and $\sigma^{2}(n)/n$ versus $n$, so as to provide simple visual comparison with the behavior of $W^{2}(a)$ functions, given in the right column. Straight lines represent the linear least-squares fits to the functions depicted.}
\label{fig2}
\end{figure*}

\begin{figure}[fig3]
%\centering
%\includegraphics[scale=0.75]{graph3.eps}
\resizebox{0.78\hsize}{!}{\includegraphics*{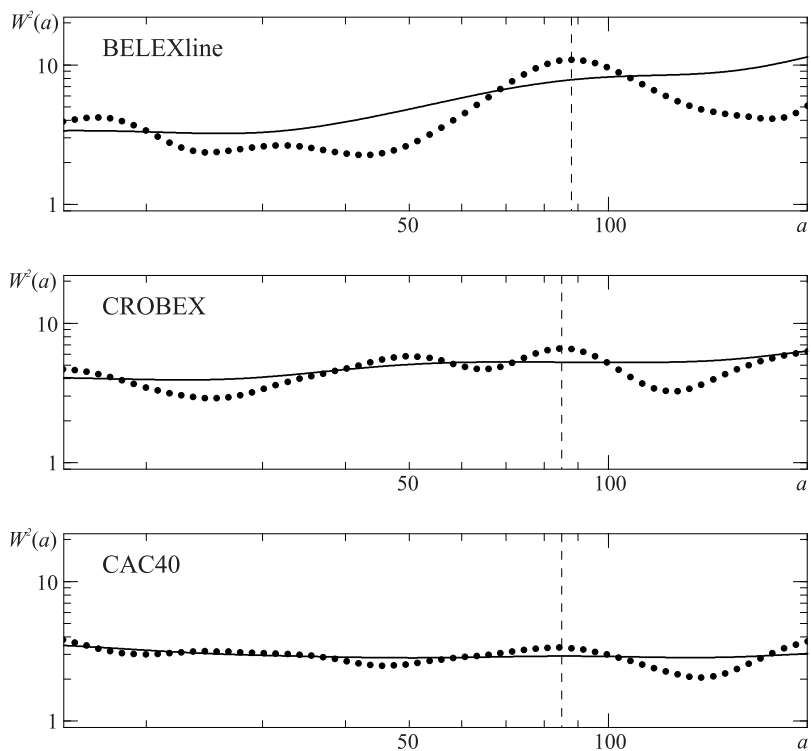}}
\caption{Results of comparison of scalegrams of the first (DOG1, solid lines) and the tenth (DOG10, small circles) order of the applied wavelet function, for the three typical examples of analyzed SMI data. It can be noticed that the application of wavelets of higher order has magnified alterations in scaling behavior of WT functions, pointing to a possible existence of external trends. The dotted vertical line marks the time period of $T_{p}= 90$ days, visible in all of the SMI series analyzed in this paper. }
\label{fig3}
\end{figure}

\begin{figure}[fig4]
%\centering
%\resizebox{0.75\textwidth}{!}{%
%  \includegraphics{graph4.eps}
%\includegraphics[scale=0.75]{graph4.eps}
\resizebox{0.78\hsize}{!}{\includegraphics*{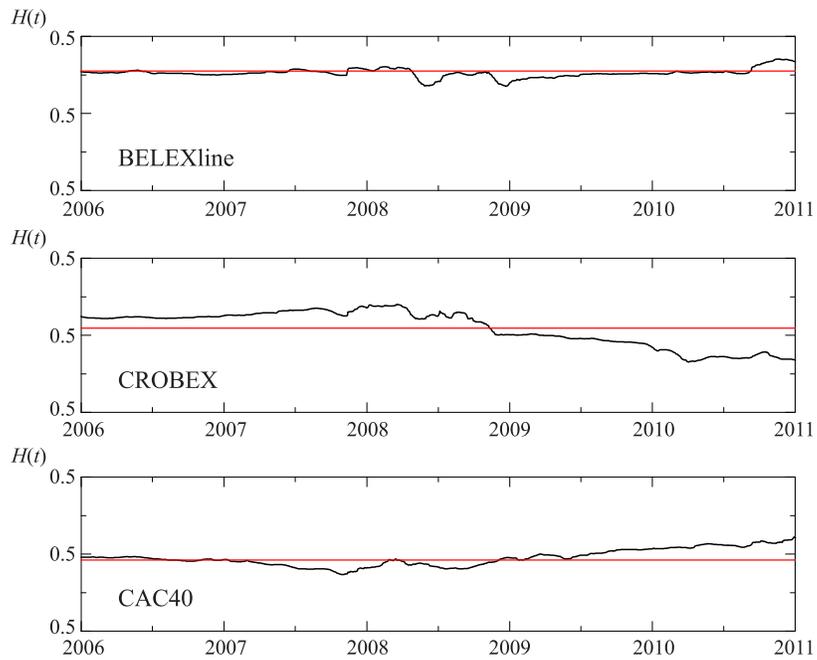}}
\caption{Local slopes of the DMA functions, giving the time-dependant values of Hurst exponent $H(t)$, as functions of time. Horizontal lines represent the calculated average (global) values of Hurst exponent $\langle H \rangle$, for three representative SMI series (BELEXline, CROBEX, and CAC40).}
\label{fig4}
\end{figure}

\end{document}